# Gravitational Equivalent Frequency and the Planck Length

*Roger Ellman*

Abstract

The mass equivalency $m \cdot c^2 = h \cdot f$ applies to gravitational mass just as to inertial mass. From that, the gravitational mass has a corresponding, associated, equivalent frequency, $f$. Using that frequency a new result is obtained in which the significance of the Planck Length, $l_{Pl}$, is clarified. The Planck Length is fundamental to gravitation and in effect supercedes $G$ in that role because it is found that there is operational or mechanical significance to the role of the Planck Length in gravitation whereas $G$ is simply a constant of proportionality.

It further is shown that the Planck Length [and, likely the Planck mass and the Planck time] are slightly mis-defined by the use of $h\text{-}bar$ [$h/2\pi$] rather than simply Planck's constant, $h$.

Theoretical implications of these results are presented. In addition, the applicability of using the frequency aspect of mass in the analysis of gravitation, and the nature of the results obtained, would appear to imply a considerably greater significance for the frequency, that is the <u>wave</u>, aspect of mass, matter, and particles in general than has been heretofore recognized.

Roger Ellman, The-Origin Foundation, Inc.
    320 Gemma Circle, Santa Rosa, CA 95404, USA
    RogerEllman@The-Origin.org

# *Gravitational Equivalent Frequency and the Planck Length*
## *Roger Ellman*

Newton's law of gravitation expressed in terms of $m_{source}$ and $m_{acted-on}$ and with both sides of the equation divided by $m_a$ is, of course,

(1) $$a_{grav} = G \cdot \frac{m_s}{d^2}$$

which states that gravitation is proportional to the mass of the gravitationally attracting body; it is a property of that body's mass.

However, mass and energy are equivalent, so that mass, $m$, is proportional to a frequency, $f$, that is characteristic of that mass. That is

(2) $$m \cdot c^2 = h \cdot f$$

$$f = \frac{c^2}{h} \cdot m$$

so that the source mass of equation *(1)*, $m_s$, has a corresponding, associated, equivalent frequency, $f_s$.

That being the case, the amount of gravitational acceleration, $a_{grav}$, can be expressed in terms of that frequency as the change, $\Delta v$, in the velocity, $v$, of the attracted mass per time period, $T_s$, of the oscillation at the corresponding frequency, $f_s$, as follows.

(3) $$a_{grav} = \Delta v / T_s = \Delta v \cdot f_s$$

It can then be reasoned as follows.

(4) $$a_{grav} = \Delta v \cdot f_s = G \cdot \frac{m_s}{d^2} \quad \text{[Equating } a_{grav} \text{ of (1) and (3)]}$$

(5) $$\Delta v \cdot \left[ \frac{m_s}{m_p} \cdot f_p \right] = G \cdot \frac{m_s}{d^2} \quad \begin{array}{l}\text{[Frequency is proportional to}\\ \text{mass and } f_p \text{ and } m_p \text{ are the}\\ \text{proton frequency and mass:}\\ f_s = (m_s/m_p) \cdot f_p.]\end{array}$$

$$\Delta v = G \cdot \frac{m_p}{d^2 \cdot f_p} \quad \text{[Rearrange, canceling } m_s \text{'s.]}$$

Then:

(6) $$\Delta v = G \cdot \frac{1}{d^2 \cdot f_p} \cdot \frac{h \cdot f_p}{c^2} \quad \text{[Substituting } m_p = h \cdot f_p/c^2\text{]}$$

$$\Delta v = G \cdot \frac{h}{d^2 \cdot c^2}$$

Now, the Planck Length, $l_{Pl}$, is defined as

(7) $$l_{Pl} \equiv \left[ \frac{h \cdot G}{2\pi \cdot c^3} \right]^{1/2} \quad \text{[the } h/2\pi \text{ part being h-bar]}$$

so that

(8) $$G = \frac{2\pi \cdot c^3 \cdot l_{Pl}^2}{h}$$

Substituting $G$ as a function of the Planck Length from equation *(8)* into $G$ as in equation *(6)*, the following is obtained.

(9) $$\Delta v = \frac{2\pi \cdot c^3 \cdot l_{Pl}^2}{h} \cdot \frac{h}{d^2 \cdot c^2}$$

$$\Delta v = c \cdot \frac{2\pi \cdot l_{Pl}^2}{d^2} \qquad \text{[Simplifying]}$$

This result states that:

- the velocity change due to gravitation, $\Delta v$,
- per cycle of the attracting mass's equivalent frequency, $f_s$,
  - which quantity, $\Delta v \cdot f_s$, is the gravitational acceleration, $a_{grav}$,
- is a specific fraction of the speed of light, $c$, namely the ratio of:
  - $2\pi$ times the Planck Length squared, $2\pi \cdot l_{Pl}^2$, to
  - the squared separation distance of the masses, $d^2$.

That squared ratio is, of course, the usual inverse square behavior.

This result also means that at distance $d = \sqrt{2\pi} \cdot l_{Pl}$ from the center of the source, attracting, mass the acceleration per cycle of that attracting mass's equivalent frequency, $f_s$, namely $\Delta v$, is equal to the full speed of light, $c$, the most that it is possible for it to be. In other words, at that [quite close] distance from the source mass the maximum possible gravitational acceleration occurs. That is the significance, the physical meaning, of $l_{Pl}$ or, rather, of $[2\pi]^{½} \cdot l_{Pl}$.

If the original definition of $l_{Pl}$ had been in terms of $h$, not $h\text{-}bar = h/2\pi$ the distinction with regard to $[2\pi]^{½}$ would not now be necessary. The $2\pi$ is a gratuitous addition, coming about from the failure to address the Hydrogen atom's stable orbits as defined by the orbital path length being an exact multiple of the orbital matter wavelength. The statement that the orbital electron's angular momentum is quantized, as in

(10) $$m \cdot v \cdot R = n \cdot \frac{h}{2\pi} \qquad [n = 1, 2, ...]$$

is merely a mis-arrangement of

(11) $$2\pi \cdot R = n \cdot \frac{h}{m \cdot v} = n \cdot \lambda_{mw} \qquad [n = 1, 2, ...]$$

the statement that the orbital path, $2\pi \cdot R$, must be an integral number of matter wavelengths, $\lambda_{mw}$, long. And, that may have resulted from a lack of confidence in the fundamental significance of matter waves because of the failure to develop theory that produced acceptable, valid, matter wave frequencies, ones such that $f_{mw} \cdot \lambda_{mw} = \textit{particle velocity}$, which is an obvious necessity. That problem is resolved in "A Reconsideration of Matter Waves"[3] where a reinterpretation of Einstein's derivation of relativistic kinetic energy (which produced his famous $E = m \cdot c^2$) leads to a valid matter wave frequency and a new understanding of particle kinetics and of the atom's stable orbits.

The physical significance of $l_{Pl}$ is in its setting of a limit on the minimum separation distance in gravitational interactions and its implying that a "core" of that radius is at the center of fundamental particles having rest mass. That is, equation *(9)* clearly implies that it is not possible for a particle having rest mass to approach another such particle closer than that distance. It is as if that distance is the radius of some impenetrable core of particles having rest mass.

That physical significance of $\sqrt{2\pi} \cdot l_{Pl}$, is so fundamental, fundamental to gravitation and apparently fundamental to particle structure, that it more truly represents a fundamental constant than does $l_{Pl}$. For those reasons that distance should replace $l_{Pl}$ as a fundamental constant of nature as follows.

*(12)* <u>The fundamental distance constant</u> $\delta$.

$$\delta^2 \equiv 2\pi \cdot l_{Pl}^2$$
$$\delta = 4.05084 \times 10^{-35} \text{ meters} \quad \text{[1986 CODATA Bulletin]}$$

Equation *(9)*, above, then becomes equation *(13)*, below,

*(13)* $\quad \Delta v = c \cdot \dfrac{\delta^2}{d^2}$

a quite pure, precise and direct statement of the operation of gravitation. It states that gravitation is a function of the speed of light, $c$, and the inverse square law, in the context of the oscillation frequency, $f_s$, corresponding to the attracting, source body's mass. It is interesting to note that equation *(13)* is exact without involving a constant of proportionality such as $G$.

There is an implication from in all of this that gravitation and the gravitational field involve something oscillatory in nature, traveling or propagating at $c$ while oscillating at $f_s$. Essentially the same description can be made of light and of all electro-magnetic radiation. It would seem somewhat absurd for material reality to involve two different, overlapping such propagations. Rather, there must be one simple such underlying form for both effects, gravitational and electro-magnetic.

There is a further implication that gravitation is directly connected to, is due to, a local slowing of $c$ in the amount of equation *(13)*. That is, gravitation being mutual between two masses; then if the attracting, source mass is propagating something toward the attracted, target mass then that latter mass is doing the same toward the former. The "local slowing" would then be the arriving propagation reducing the encountered target's own propagation.

And, such an effect could, and should, require some adjustment by the target so as to maintain the flow of its propagation at the un-slowed value of $c$. Such an adjustment would be a compensating increase in the target's velocity toward the source, namely the already obtained *Δv*.

Such slowing correlates with gravitational lensing's light path bending. If propagating light waves are the same "underlying form" [posited in the third preceding paragraph] as the gravitational propagation being discussed, which slows an encountered similar flow, then light propagation passing a gravitating mass would experience greater slowing of its wave front on the portion nearer to the gravitating mass and lesser slowing further away [because of the inverse square behavior] - effects tending to bend the direction of the wave front somewhat toward the attracting mass as in the observed gravitational lensing.

The applicability of using the frequency aspect of mass in the analysis of gravitation, and the nature of the results obtained, would appear to imply a considerably greater significance for the frequency, that is the wave, aspect of mass, matter, and particles in general than has been heretofore recognized.